%% file: main.tex
\documentclass[reprint,showkeys,longbibliography,nofootinbib,superscriptaddress,aps,notitlepage]{revtex4-2}

\usepackage[utf8]{inputenc}
\usepackage{mathtools}
\usepackage{amsmath}
\usepackage{amssymb}
\usepackage[colorlinks,unicode]{hyperref}
\usepackage{graphicx}
\usepackage{multirow}
\usepackage{physics}
\usepackage{siunitx}
\usepackage{xspace} 
\usepackage{lineno}

\newcommand{\geant}{\textsc{Geant4}\xspace}
\newcommand{\elec}{$\mathrm{e^{-}}$}
\newcommand{\Am}{$\mathrm{^{241}Am}$ }

\begin{document}
\preprint{}

\title{Precision measurement of Compton scattering in silicon with a skipper CCD for dark matter detection}
\input{0_authorlist}

\date{\today} 

\begin{abstract}
\noindent Experiments aiming to directly detect dark matter through particle recoils can achieve energy thresholds of $\mathcal{O}(10\,\mathrm{eV})$. In this regime, ionization signals from small-angle Compton scatters of environmental $\gamma$-rays constitute a significant background. Monte Carlo simulations used to build background models have not been experimentally validated at these low energies. We report a precision measurement of Compton scattering on silicon atomic shell electrons down to 23\,eV. A skipper charge-coupled device (CCD) with single-electron resolution, developed for the DAMIC-M experiment, was exposed to a \Am{} $\gamma$-ray source over several months. Features associated with the silicon K, L$_{1}$, and L$_{2,3}$-shells are clearly identified, and scattering on valence electrons is detected for the first time below 100\,eV. We find that the relativistic impulse approximation for Compton scattering, which is implemented in Monte Carlo simulations commonly used by direct detection experiments, does not reproduce the measured spectrum below 0.5\,keV. The data are in better agreement with {\it ab initio} calculations originally developed for X-ray absorption spectroscopy.
\end{abstract}
\keywords{dark matter, charge-coupled devices, silicon detectors, Compton scattering, NRIXS}

\maketitle

\input{1_introduction.tex}
\input{2_compton.tex}
\input{3_setup.tex}
\input{4_reconstruction.tex}

\input{4p_simulation.tex}
\input{5_analysis.tex}

\input{6_model.tex}
\input{7_conclusion.tex}
\input{8_acknowledgements}

\bibliographystyle{style/h-physrev}
\bibliography{compton.bib}

\end{document}

%% file: 0_authorlist.tex

\author{D.\,Norcini}\email[Corresponding author: ]{dnorcini@kicp.uchicago.edu}
\affiliation{Kavli Institute for Cosmological Physics and The Enrico Fermi Institute, The University of Chicago, Chicago, IL, United States}
\author{N.\,Castell\'{o}-Mor}
\affiliation{Instituto de F\'{i}sica de Cantabria (IFCA), CSIC - Universidad de Cantabria, Santander, Spain}

\author{D.\,Baxter}\email[Now at Fermi National Accelerator Laboratory, Batavia, IL, United States]{}
\author{N.J.\,Corso}
\author{J.\,Cuevas-Zepeda}
\affiliation{Kavli Institute for Cosmological Physics and The Enrico Fermi Institute, The University of Chicago, Chicago, IL, United States}
\author{C.\,De Dominicis}
\affiliation{SUBATECH, Nantes Universit\'{e}, IMT Atlantique, CNRS-IN2P3, Nantes, France}
\affiliation{Kavli Institute for Cosmological Physics and The Enrico Fermi Institute, The University of Chicago, Chicago, IL, United States}
\author{A.\,Matalon}
\affiliation{Kavli Institute for Cosmological Physics and The Enrico Fermi Institute, The University of Chicago, Chicago, IL, United States}
\affiliation{Laboratoire de physique nucl\'{e}aire et des hautes \'{e}nergies (LPNHE), Sorbonne Universit\'{e}, Universit\'{e} Paris Cit\'{e}, CNRS/IN2P3, Paris, France}
\author{S.\,Munagavalasa}
\author{S.\,Paul}
\affiliation{Kavli Institute for Cosmological Physics and The Enrico Fermi Institute, The University of Chicago, Chicago, IL, United States}
\author{P.\,Privitera}
\affiliation{Kavli Institute for Cosmological Physics and The Enrico Fermi Institute, The University of Chicago, Chicago, IL, United States}
\affiliation{Laboratoire de physique nucl\'{e}aire et des hautes \'{e}nergies (LPNHE), Sorbonne Universit\'{e}, Universit\'{e} Paris Cit\'{e}, CNRS/IN2P3, Paris, France}
\author{K.\,Ramanathan}\email[Now at the California Institute of Technology, Pasadena, CA, United States]{}
\author{R.\,Smida}
\author{R.\,Thomas}
\author{R.\,Yajur}
\affiliation{Kavli Institute for Cosmological Physics and The Enrico Fermi Institute, The University of Chicago, Chicago, IL, United States}

\author{A.E.\,Chavarria}
\author{K.\,McGuire}
\author{P.\,Mitra}
\author{A.\,Piers}
\affiliation{Center for Experimental Nuclear Physics and Astrophysics, University of Washington, Seattle, WA, United States}

\author{M.\,Settimo}
\affiliation{SUBATECH, Nantes Universit\'{e}, IMT Atlantique, CNRS-IN2P3, Nantes, France}

\author{J.\,Cortabitarte Guti\'{e}rrez}
\author{J.\,Duarte-Campderros}
\author{A.\,Lantero-Barreda}
\author{A.\,Lopez-Virto}
\author{I.\,Vila}
\author{R.\,Vilar}
\affiliation{Instituto de F\'{i}sica de Cantabria (IFCA), CSIC - Universidad de Cantabria, Santander, Spain}

\author{N.\,Avalos}
\affiliation{Centro At\'{o}mico Bariloche and Instituto Balseiro, Comisi\'{o}n Nacional de Energ\'{i}a At\'{o}mica (CNEA), Consejo Nacional de Investigaciones Cient\'{i}ficas y T\'{e}cnicas (CONICET), Universidad Nacional de Cuyo (UNCUYO), San Carlos de Bariloche, Argentina}
\author{X.\,Bertou}
\affiliation{Centro At\'{o}mico Bariloche and Instituto Balseiro, Comisi\'{o}n Nacional de Energ\'{i}a At\'{o}mica (CNEA), Consejo Nacional de Investigaciones Cient\'{i}ficas y T\'{e}cnicas (CONICET), Universidad Nacional de Cuyo (UNCUYO), San Carlos de Bariloche, Argentina}
\author{A.\,Dastgheibi-Fard}
\affiliation{LPSC LSM, CNRS/IN2P3, Universit\'{e} Grenoble-Alpes, Grenoble, France}
\author{O.\,Deligny}
\affiliation{CNRS/IN2P3, IJCLab, Universit\'{e} Paris-Saclay, Orsay, France}
\author{E.\,Estrada}
\affiliation{Centro At\'{o}mico Bariloche and Instituto Balseiro, Comisi\'{o}n Nacional de Energ\'{i}a At\'{o}mica (CNEA), Consejo Nacional de Investigaciones Cient\'{i}ficas y T\'{e}cnicas (CONICET), Universidad Nacional de Cuyo (UNCUYO), San Carlos de Bariloche, Argentina}
\author{N.\,Gadloa}
\affiliation{Universit\"{a}t Z\"{u}rich Physik Institut, Z\"{u}rich, Switzerland}
\author{R.\,Ga\"{i}or}
\affiliation{Laboratoire de physique nucl\'{e}aire et des hautes \'{e}nergies (LPNHE), Sorbonne Universit\'{e}, Universit\'{e} Paris Cit\'{e}, CNRS/IN2P3, Paris, France}
\author{T.\,Hossbach}
\affiliation{Pacific Northwest National Laboratory (PNNL), Richland, WA, United States} 
\author{L.\,Khalil}
\affiliation{Laboratoire de physique nucl\'{e}aire et des hautes \'{e}nergies (LPNHE), Sorbonne Universit\'{e}, Universit\'{e} Paris Cit\'{e}, CNRS/IN2P3, Paris, France}
\author{B.\,Kilminster}
\affiliation{Universit\"{a}t Z\"{u}rich Physik Institut, Z\"{u}rich, Switzerland}
\author{I.\,Lawson}
\affiliation{SNOLAB, Lively, ON, Canada }
\author{S.\,Lee}
\affiliation{Universit\"{a}t Z\"{u}rich Physik Institut, Z\"{u}rich, Switzerland}
\author{A.\,Letessier-Selvon}
\affiliation{Laboratoire de physique nucl\'{e}aire et des hautes \'{e}nergies (LPNHE), Sorbonne Universit\'{e}, Universit\'{e} Paris Cit\'{e}, CNRS/IN2P3, Paris, France}
\author{P.\,Loaiza}
\affiliation{CNRS/IN2P3, IJCLab, Universit\'{e} Paris-Saclay, Orsay, France}
\author{G.\,Papadopoulos}
\affiliation{Laboratoire de physique nucl\'{e}aire et des hautes \'{e}nergies (LPNHE), Sorbonne Universit\'{e}, Universit\'{e} Paris Cit\'{e}, CNRS/IN2P3, Paris, France}
\author{P.\,Robmann}
\affiliation{Universit\"{a}t Z\"{u}rich Physik Institut, Z\"{u}rich, Switzerland}
\author{M.\,Traina}
\affiliation{Laboratoire de physique nucl\'{e}aire et des hautes \'{e}nergies (LPNHE), Sorbonne Universit\'{e}, Universit\'{e} Paris Cit\'{e}, CNRS/IN2P3, Paris, France}
\author{G.\,Warot}
\affiliation{LPSC LSM, CNRS/IN2P3, Universit\'{e} Grenoble-Alpes, Grenoble, France}
\author{J-P.\,Zopounidis}
\affiliation{Laboratoire de physique nucl\'{e}aire et des hautes \'{e}nergies (LPNHE), Sorbonne Universit\'{e}, Universit\'{e} Paris Cit\'{e}, CNRS/IN2P3, Paris, France}

\collaboration{DAMIC-M Collaboration}

%% file: 1_introduction.tex

\section{Introduction} 
\label{sec:introduction}
The existence of dark matter, a preponderant, non-luminous material that interacts gravitationally, has been unequivocally established by astrophysical and cosmological observations\,\cite{ParticleDataGroup:2020ssz}.
The hypothesis that dark matter is made of unknown particles has compelled experimental searches to directly detect them through interactions with target materials\,\cite{ParticleDataGroup:2020ssz}.
Many of these experiments are located in underground laboratories to shield from cosmic-rays and need further sophisticated techniques to suppress radiogenic and cosmogenic backgrounds\,\cite{doi:10.1146/annurev.ns.45.120195.002551,Mei:2005,10.3389/fphy.2020.577734}. Still, energy deposits from Compton scattered $\gamma$-rays constitute a significant background in detectors searching for dark matter particles. Compton scattering may occur deep in the active detection volume mimicking a dark matter interaction. While in some experiments nuclear recoils induced by weakly interacting massive particles (WIMPs)\,\cite{Kolb:1990vq} have a signature distinct from that of Compton ionization signals, the rejection power drastically decreases at low energy\,\cite{PhysRevLett.112.241302,CRESST:2020wtj,PhysRevD.102.112002}. Dark matter–electron interactions expected in the so-called dark sector models\,\cite{Alexander:2211864,https://doi.org/10.48550/arxiv.1311.0029} are indistinguishable from Compton scattering. A precise knowledge of the Compton background spectrum in the detector material is thus of paramount importance in finding evidence of the interactions of dark matter particles.

The DAMIC-M (Dark Matter in CCDs at Modane) experiment employs skipper charge-coupled devices (CCDs) to directly search for the interactions of dark matter particles\,\cite{damicm2020}. Compared to scientific CCDs with conventional readout, e.g. those used by the precursor DAMIC experiment\,\cite{PhysRevD.94.082006,DAMIC:2019dcn,PhysRevLett.125.241803,Aguilar_Arevalo_2021,PhysRevD.105.062003}, skipper CCDs allow for a significant reduction in readout noise, enabling the detection of single electrons and energy thresholds of a few eV\,\cite{skipper,Chandler1990zz,Tiffenberg:2017aac}. With skipper readout, DAMIC-M has unprecedented sensitivity to nuclear and electronic recoils from interactions of low-mass (1-10$^4$\,MeV/c$^2$) dark matter candidates in the bulk silicon of the CCDs. Understanding backgrounds down to the DAMIC-M energy threshold is essential for exploiting this sensitivity.

Small-angle Compton scatters produce low-energy electron recoils, including from freed atomic shell electrons which have a well-defined binding energy. Ionization signals from shell electrons should produce a spectrum with predicted features according to their binding energy in the region below 200\,eV, which corresponds to L-shell transition energies in silicon. Compton scattering produces point-like energy deposits uniformly distributed in the silicon bulk, as is the case for dark matter particles interactions. A precise measurement of the Compton spectrum will allow the detector response to be calibrated down to a few electrons, improving sensitivity for low-mass dark matter searches.

In this paper, we report a precision measurement of Compton scattering on silicon shell electrons in a skipper CCD. Measurements were performed by exposing a DAMIC-M prototype CCD to a \Am{} $\gamma$-ray source. Our result improves upon previous work using scientific CCDs with conventional readout\,\cite{Ramanathan2017MeasurementDetector} and an energy threshold of 60\,eV, where an unexpected softening of the spectrum in the L-shell region was observed. Results from another group also using a skipper CCD observed a similar effect\,\cite{Botti:2022lkm}. Here we present measurements with sub-electron charge resolution down to 23\,eV, allowing for a robust mapping of the spectrum in a region that has not been previously measured.

%% file: 2_compton.tex

\section{Compton scattering} 
\label{sec:compton}

Compton scattering describes the interaction between an incident photon and a free electron\,\cite{PhysRev.21.483}. Such interactions result in the deflection of the photon and recoil of the electron, whose energy and direction can be obtained by conservation laws. 
Assuming a free electron at rest, the interaction cross section is described by the well-known Klein-Nishina formula\,\cite{Klein}. For the scattering of unpolarized photons with an atomic electron in a target, the double-differential cross section may be more generally expressed as\,\cite{H_m_l_inen_2001}:

\begin{equation}
\left. \frac{d^{2}\sigma}{dE \cdot d\Omega} \right| _{nl}= r_{0}^2 \left( \frac{1+  \cos^{2} \theta}{2} \right) \left ( 1 - \frac{E}{E_{\gamma}} \right ) S_{nl}(q,E),
\label{eq:genxsection}
\end{equation}

\noindent where $r_{0}$ is the classical electron radius, $\theta$ is the scattering angle of the deflected photon, $E_{\gamma}$ is the initial photon energy, $E$ is the difference between the initial and final photon energies (i.e. the energy deposited in the target), and $q$ is the magnitude of the photon scattering vector. The dynamic structure factor $S_{nl}(q,E)$ encapsulates the target-dependent component of the cross section and depends on the atomic quantum numbers $n$ and $l$ of the target electron.

The simplest extension of the Klein-Nishina formula is to treat each atomic shell electron as free but with a constrained momentum distribution. This is achieved in the so called relativistic impulse approximation (RIA)\cite{PhysRevA.26.3325}, where the dynamic structure factor is expressed as

\begin{equation}
S_{nl}(q,E) = \frac{m}{q}\chi(p_z)J_{nl}(p_z),
\label{eq:SJeq}
\end{equation}

\noindent where $m$ is the mass of the electron and $\chi(p_z)$ is the relativistic correction factor. The Compton profile $J_{nl}$ depends only on $p_z$, the projection of the initial electron momentum on the photon scattering vector. Tables of computed $J_{nl}$ for different atomic electrons can be found in the literature\,\cite{BIGGS1975201}. The RIA succeeds in describing several features of the deposited-energy spectrum, including the broadening of the Compton edge.

\begin{figure}[h!]
    \centering
    \includegraphics[width=\linewidth]{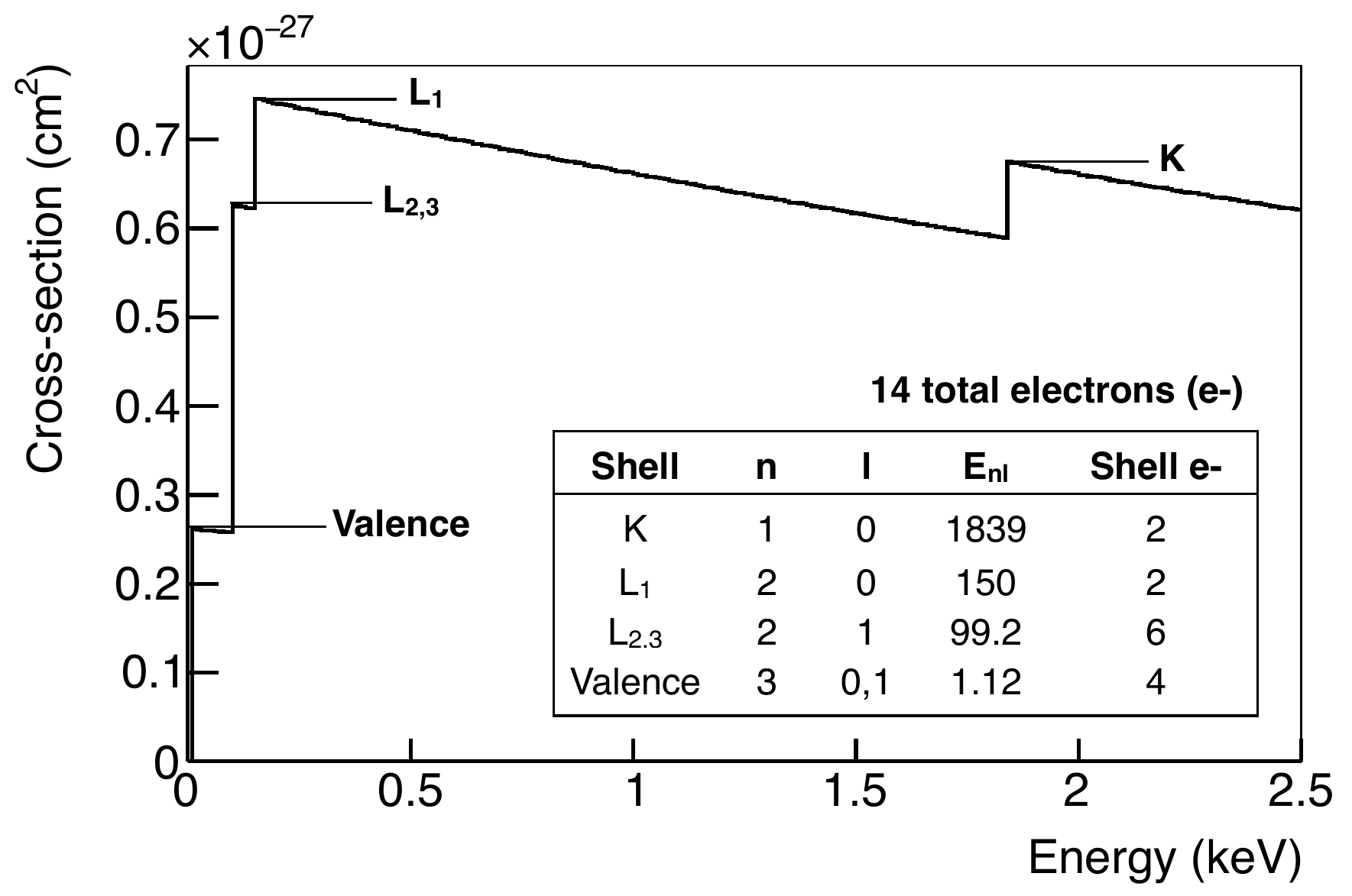}
    \caption{Electron spectrum calculated using the RIA approximation for Compton scattering of a 59.54\,keV photon in silicon. The inset table details the quantum numbers $(n,l)$, binding energy ($E_{nl}$), and number of electrons in each atomic shell. Each shell level is labeled on the spectrum. The relative height of each step is approximately the ratio of the electrons in each shell to the total available electrons.}
   \label{fig:iamodel_spectrum}
\end{figure}

A low-energy prediction from the RIA is constructed by considering the atomic binding energies $E_{nl}$ in relation to the transfer energy.
When $E\,<\,E_{nl}$, i.e. the energy transfer is less than the binding energy of the atomic shell, $d\sigma/dE d\Omega|_{nl}=$\,0 and the energy distribution forms steps proportional to the number of electrons in the shell. At $E=E_{nl}$, the scattered electron has negligible kinetic energy and the photon is likely to escape after a single scatter, especially in a thin detector such as the CCD used in this measurement. Deposited energy is thus a result of refilling the atomic vacancy by emission of secondary Auger/Coster–Kronig\,\cite{COSTER193513} electrons or fluorescence X-rays. For $E\,>\,E_{nl}$, freed electron energies fall on an approximately constant slope between steps obtained from the integration of Equation\,\ref{eq:genxsection} over all scattering angles (momentum transfers).
The resulting spectrum is shown in Figure\,\ref{fig:iamodel_spectrum} for incident \Am $\gamma$-rays ($E_\gamma=$\,59.5\,keV) on silicon. A similar procedure\,\cite{BRUSA1996167} is implemented in the particle-tracking Monte Carlo code \geant\,\cite{AGOSTINELLI2003250}, widely used for the estimate of backgrounds in experiments directly searching for dark matter interactions.

The approximation of a free electron may not be adequate in the regime where the energy transfer is comparable to the electron binding and kinetic energies\,\cite{PRATT2007175}.
In this case, it is more appropriate to compute $S_{nl}(q,E)$ with \emph{ab initio} calculations. 
We use the FEFF\,\cite{Kas:hf5422, feff} code, which performs a full quantum mechanical treatment to sum over all transition probabilities from the initial state to all possible atomic final states in the target material.
FEFF was primarily developed (and has been extensively validated) for X-ray absorption spectroscopy\,\cite{RevModPhys.72.621} but includes the option to calculate a $S_{nl}(q,E)$ for non-resonant inelastic X-ray scattering (NRIXS)\,\cite{PhysRevB.72.045136}\footnote{The process referred to as ``Compton scattering" in this paper is more commonly known as NRIXS in the X-ray community.}.

The previous measurement performed with a silicon CCD\,\cite{Ramanathan2017MeasurementDetector} found the RIA model to reasonably describe the data in the K-shell region. However, the measured spectrum in the L-shells region was notably softer than the model prediction, with deviations that could not be accounted for by the resolution of the experiment. Separate L$_1$ and L$_{2,3}$ step features were not observed.
In this paper, we refine the measurement of the deposited-energy spectrum from 59.5\,keV $\gamma$-rays scattering in silicon, and compare to the predictions from RIA and FEFF.

%% file: 3_setup.tex

\section{Experimental setup and data sets} 
\label{sec:setup}
\begin{figure}[h!]
    \centering
    \includegraphics[width=0.98\linewidth]{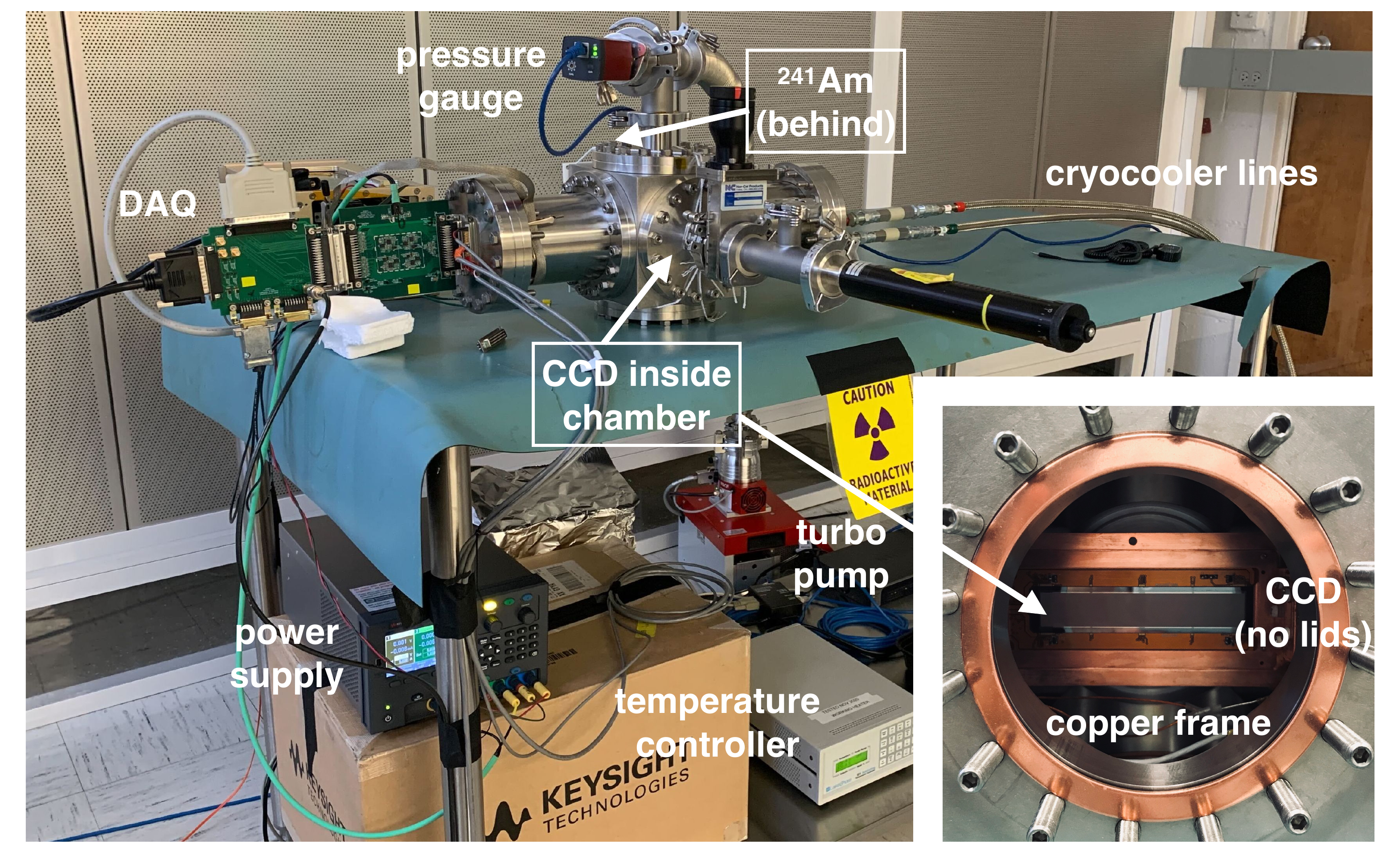}
    \caption{Experimental setup used for the Compton scattering measurement. The 1024$\times$6176 skipper CCD is mounted in a copper frame, as shown in the inset (lids not pictured), inside of the vacuum chamber. Components required for operation (e.g. the turbo pump, cryocooler, and electronics) interface with the chamber via flanges. The \Am source is positioned behind the chamber, centered with the CCD plane.}
    \label{fig:Setup}
\end{figure}

The experimental setup for this measurement is located in an on-surface clean room at The University of Chicago, as shown in Figure\,\ref{fig:Setup}.
A skipper CCD with 1024$\times$6176 pixels is used as the silicon target and detector. It features a three-phase polysilicon gate structure with a buried p-channel, pixel size 15$\times$15\,$\mathrm{\mu m^{2}}$, and a thickness of 675\,$\mathrm{\mu m}$. The bulk of the device is high-resistivity (10–20 k$\Omega$\,cm) n-type silicon which allows for fully depleted operation at substrate biases $\ge$\,40\,V. The CCD was developed at Lawrence Berkeley National Laboratory MicroSystems Lab\,\cite{Holland:2002zz,Holland:2003zz,Holland:2009zz} and fabricated by Teledyne DALSA Semiconductor as a prototype for the DAMIC-M experiment. Wirebonding and packaging was completed at the University of Washington.

The CCD is mounted in a copper frame within a stainless-steel vacuum chamber held at a pressure of $10^{-7}$\,mbar and cooled to 126\,K. Thin 1.6\,mm aluminum lids are placed on both the front and backside of the copper frame to shield the CCD from IR photons generated by the warm chamber walls. A \Am source (59.54\,keV $\gamma$-ray) is mounted on the chamber illuminating the backside of the CCD, shielded by a 1.3\,cm aluminum block to suppress weak lines between $\sim$10-35\,keV. This source was chosen because it has an intense $\gamma$-line at an energy where Compton scattering is the dominant interaction with resulting electron recoils fully depositing their energy in the bulk. The voltage biases, clocks, and video signals required for the CCD operation are provided by a Kapton flex cable wirebonded to the device. The silicon bulk is kept fully depleted by the application of a 95\,V external bias. The CCD is controlled and read out by a custom data acquisition system based on commercial CCD electronics from Astronomical Research Cameras, Inc. A slow control system is used to operate the various instruments and monitor their status\,\cite{Nikkel_ASC,Norcini_Chicago_Slow_Control_2020}.

Within the CCD, Compton scattering generates charge carriers in the bulk silicon that are proportional to the energy deposited by the interaction. The voltage bias applied between the bottom and top surface of the device drifts the charge along the z-direction towards the pixel array. Charge also diffuses in the lateral directions due to thermal motion, with a spatial spread $\mathrm{\sigma_{xy}}$ proportional to the drift length. Thus, clusters of pixels with charge found in the CCD images identify the location of interactions both in the xy-plane and the z-direction\,\cite{PhysRevD.94.082006}. 

\begin{figure}[h!]
    \centering  
    \includegraphics[width=\linewidth]{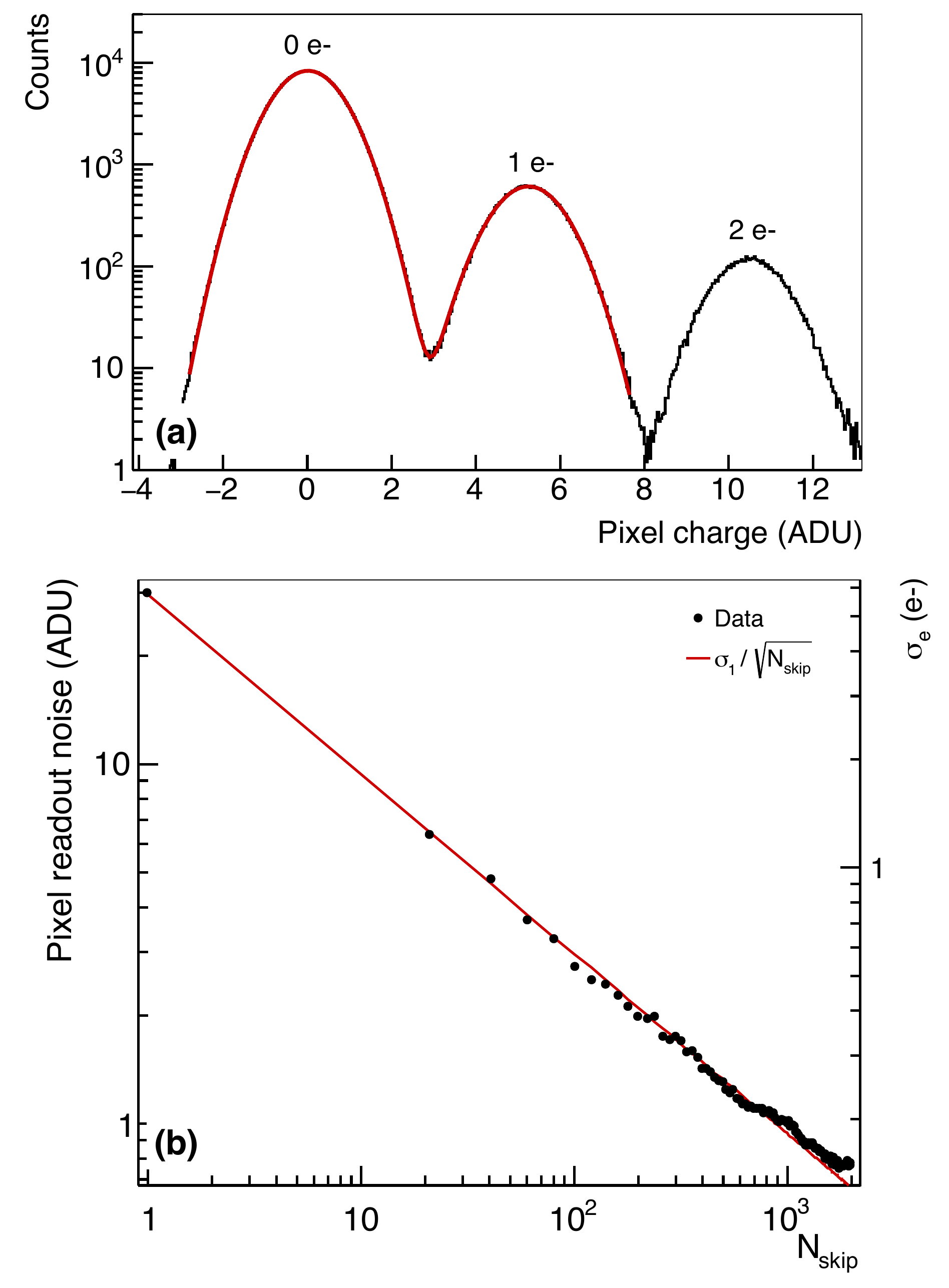}
    \caption{(a) Pixel charge distribution, where the pixel charge is obtained from the average of $N_{skip}=$\,2000 NDCMs. Individual peaks correspond to 0, 1, and 2 electrons. From the distance between fitted peaks (red) a calibration constant of 5.1\,ADU/\elec{} is derived (in ADC Units, ADU). The pixel readout noise, obtained from the standard deviation of the zero electron distribution, is $\sigma_{\text{e}}=$\,0.13\,\elec. (b) Pixel readout noise as a function of $N_{skip}$. The data scales with the $1/\sqrt{N_{skip}}$ expectation (red) for independent, uncorrelated measurements.}
    \label{fig:skips}
\end{figure}

The voltage clocks move the charge held in a pixel row-by-row towards the serial register of the CCD. The charge is then clocked to the end of the serial register where a charge-to-voltage amplifier is located for readout. Unlike conventional CCDs, skipper CCDs\,\cite{skipper,Chandler1990zz,Tiffenberg:2017aac} can be configured to make multiple non-destructive charge measurements (NDCMs). Skipper readout essentially moves the charge contained in each pixel back-and-forth into the readout node allowing for many measurements of the same pixel, so that they can be averaged. Since the measurements are uncorrelated, the readout noise is then reduced to $\sigma_{N_{skip}}=\sigma_{1}/\sqrt{N_{skip}}$, where $\sigma_{1}$ is the single-sample readout noise (the standard deviation of a single charge measurement) and $N_{skip}$ is the number of NDCMs. By taking a large enough number of NCDMs, the readout noise can reach the sub-electron level and the detection threshold is reduced accordingly. Figure\,\ref{fig:skips} demonstrates the achieved noise of the skipper readout allowing single-electron\footnote{For the sake of simplicity, we use the term electrons to indicate charge carriers detected in the CCD. However, holes are held in the pixels of the p-channel CCD used for this measurement.} charges to be resolved. The single-electron resolution also provides a straightforward way of calibrating the energy response of the detector (see Section\,\ref{sec:recon}). 

Data collection is automated and taken on a run-by-run basis. Each run consists of multiple image types where the image size, number of NDCMs, and pixel binning\footnote{Pixel binning is an operating mode of the CCD where the charge of several pixels is summed before being read out. A n$\times$m binning corresponds to summing the charge of n pixels in the serial direction and m pixels in the parallel direction.} are varied. Full CCD images with no binning and $N_{skip}=$\,1 are taken to monitor the overall quality of the device, including stability of defects (faulty pixels) over the active area. Images with $N_{skip}=$\,2000 are taken to calibrate the energy response and dark current of the CCD with maximum resolution. For the analysis presented here, the CCD operating parameters are optimized to ensure good resolution while reducing occupancy to avoid overlapping clusters that can bias the measured energy spectrum. Each image corresponds to 10\% of the CCD active area. A 4$\times$4 binning is used to collect all charge from a single interaction into fewer pixels, thus reducing the contribution of the readout noise to the charge measurement. Every binned pixel is read out with $N_{skip}=$\,64, further reducing the noise. Remaining charge on the CCD is cleared before each image to guarantee that all images have the same exposure. A background data set without the \Am source is taken with the same parameters as the Compton analysis data. Another data set with the \Am source and parallel clocks moving charge towards the active region allowed the study of backgrounds within the serial register (see Section\,\ref{sec:analysis} for details).

Table\,\ref{tab:datasets} summarizes the main data sets and details the relevant CCD operating parameters. Image sizes in the parallel direction that exceed the active area of the CCD are known as the ``overscan". These additional overscan pixels do not contain charge and are used to determine the baseline for each row. 
Note that in the subsequent sections, a ``pixel" refers to a 4$\times$4 binned pixel for $N_{skip}=$\,64 data and to a 16$\times$4 binned pixel for $N_{skip}=$\,2000 data.

\begin{table*}[!ht]
    \centering
    \begin{tabular}{ccccccccccccc}\hline\hline
        Source  &Binning           & Image size        & NDCMs & Charge direction & Images    & Exposure & Cluster density & Type            \\
                &[row$\times$col]  & [row$\times$col]  &       &        &           & [days]   & [evt/keV/image] &                 \\\hline
        \Am     & 1$\times$1       & 6200$\times$1100  & 1     & serial register    & 909       & 1.3      & --              & diagnostic      \\
                & 16$\times$4      & 60$\times$275     & 2000  & serial register   & 909       & 5.9      & --              & calibration     \\    
                & 4$\times$4       & 150$\times$275    & 64    & serial register   & 223\,517   & 105.5    & 3.32            & source          \\
                & 4$\times$4       & 150$\times$275    & 64    & active area    & 26\,948    & 11.8     & 0.07            & serial register \\
        None    & 4$\times$4       & 150$\times$275    & 64    & serial register   & 103\,106   & 48.1     & 0.02            & background      \\\hline \hline 
    \end{tabular}
    \caption{Summary of the data sets used in the analysis. Overscan columns that exceed the active area of the CCD. The charge direction parameter represents the direction that charge is moved, either towards the readout serial register or away into the active area of the CCD. The cluster density was estimated in the 1-5\,keV range and includes only events that pass selection cuts. All data were taken with a substrate voltage of 95\,V to limit the lateral diffusion allowing charge from a single interaction to be collected in 4$\times$4 binned pixels.}
    \label{tab:datasets}
\end{table*}

Data collection lasted several months requiring continuous monitoring of the data quality. Automatic data analysis reports for each run provide information on the pedestal baselines, dark current level, calibration constants, readout noise and single-electron resolution, allowing for an accurate tracking of the CCD performance and stability. All monitored parameters show excellent stability to a few per mil, with less than 1\% of the data rejected due to occasional spurious noise sources in the laboratory. 

%% file: 4_reconstruction.tex

\section{Image processing, calibration and event reconstruction}
\label{sec:recon}

\begin{figure}[h!]
    \centering
    \includegraphics[width=\linewidth]{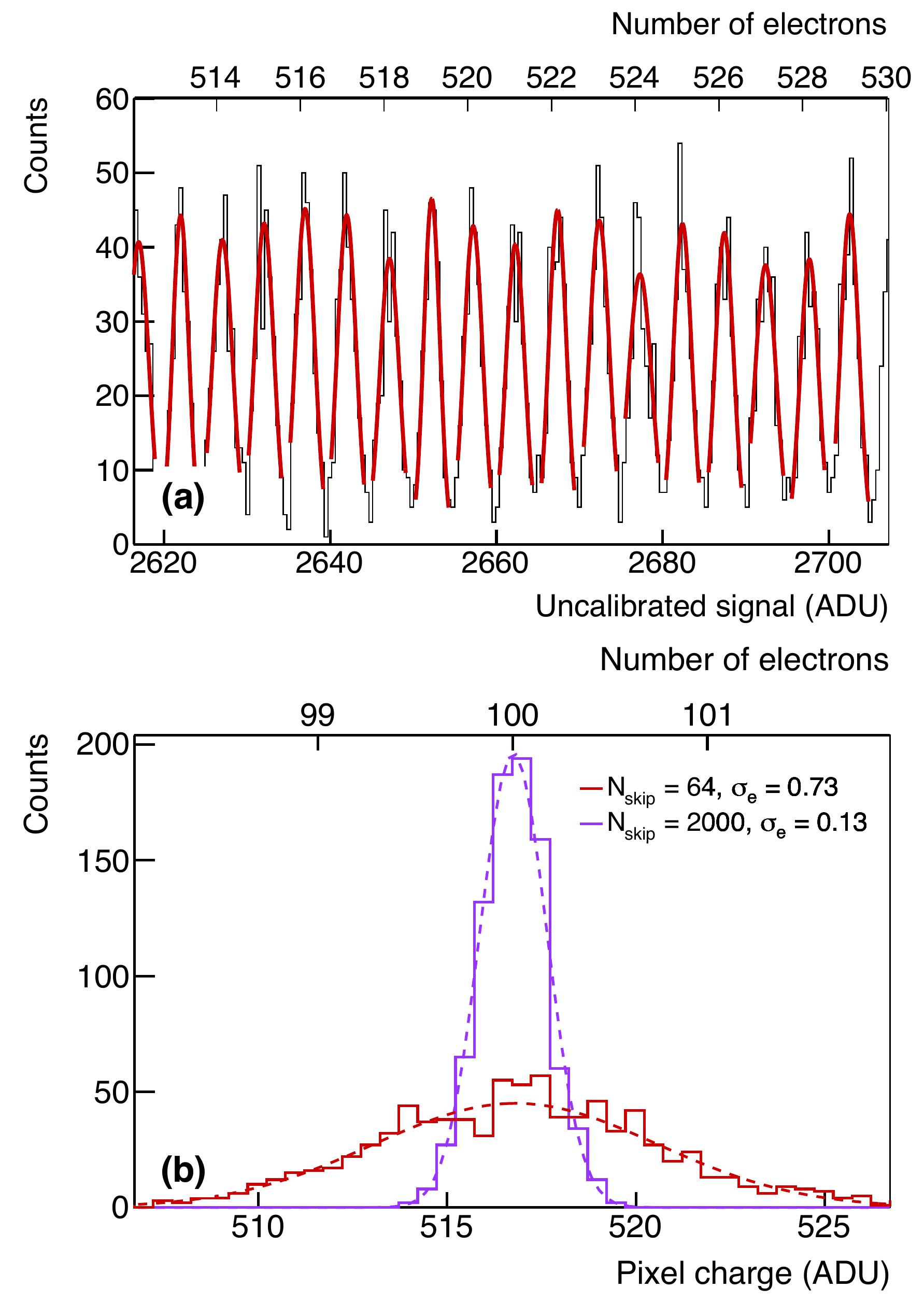}
    \caption{(a) Pixel charge distribution with single-electron resolution ($N_{skip}=$\,2000) showing individual peaks up to 550 electrons ($\sim$2.1\,keV). To calibrate the energy scale, the mean value of each peak in ADU is compared with the corresponding number of electrons. (b) Pixel charge distribution with $N_{skip}=$\,2000 (purple) and $N_{skip}=$\,64 (red) corresponding to a charge of 100 electrons. The pixels entering the two distributions are the same, with their charge calculated using only the first 64 out of the 2000\,NCDMs for $N_{skip}=$\,64.}
    \label{fig:calibration}
\end{figure}

\begin{figure}[h!]
    \centering
    \includegraphics[width=\linewidth]{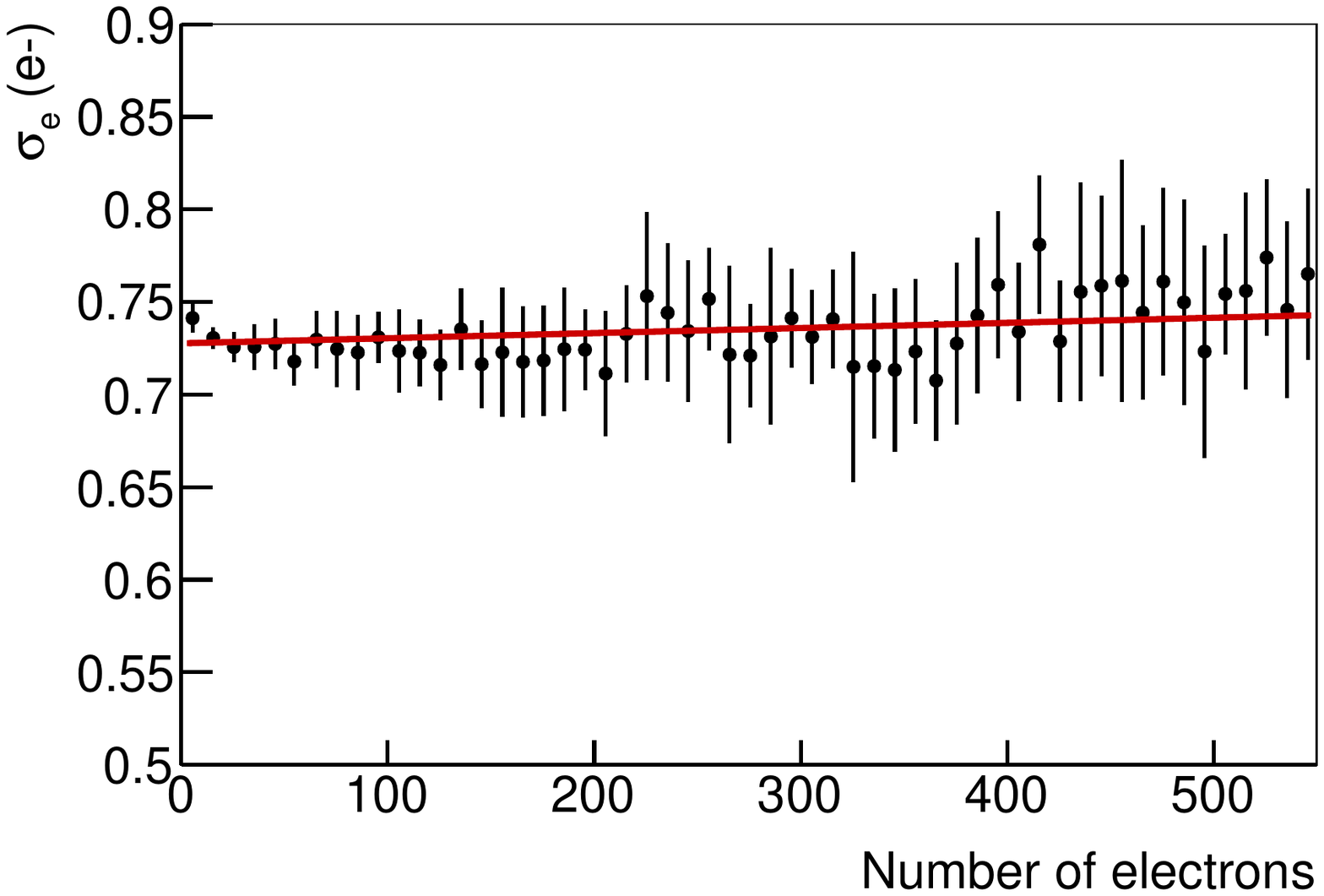}
    \caption{Charge resolution for $N_{skip}=$\,64 as a function of charge up to 550\,\elec{} ($\sim$2.1\,keV). The resolution changes by $<$\,3\% over the entire energy range (red).}
    \label{fig:64_res}
\end{figure}

A raw image contains an array of signals (in ADC Units, ADU) corresponding to every charge measurement performed by the readout chain. Image processing begins by calculating the average signal using the $N_{skip}$ values for each pixel. To eliminate the DC offset of the electronics chain, a pedestal value is obtained from a Gaussian fit of the pixel value distribution in the overscan region, which has mostly zero charge. The pedestal is determined independently for each row and then subtracted from all pixels in the row. At the end of this procedure, the resulting pixel value in ADU is proportional to the charge contained in the pixel. We then exploit the single-electron resolution provided by the skipper readout to determine the calibration between ADU and electrons, using $N_{skip}=$\,2000 images with a resolution of $\sigma_{\text{e}}=$\,0.13\,\elec{} (where the single-sample noise is $\sigma_{1}=$6\,\elec{}). The pixel charge distribution obtained from these images shows over 550 consecutive peaks individually resolved with sufficient statistical precision (see Figure\,\ref{fig:calibration}{\color{red}(a)}), where the number of $k-1$ electrons corresponds to peak $k$ (zero electrons are associated to the first peak). Since the Compton measurement is performed with $N_{skip}=$\,64, a further step is required, as illustrated in Figure\,\ref{fig:calibration}{\color{red}(b)}): for each peak, the value of the associated pixels is recalculated using only the first 64 out of the 2000 NCDMs. The mean value and standard deviation in ADU of this pixel distribution with $N_{skip}=$\,64 is obtained from a Gaussian fit. The calibration is then performed by comparing the mean value in ADU to the number of electrons corresponding to the peak. With this procedure the charge linearity is measured up to $550$\,\elec{} ($\sim$2.1\,keV), covering the entire region of interest, and found to be stable within 3\% throughout. For a precise conversion of ADU to electrons, which takes into account residual non-linearity, we use a two-degree polynomial. This procedure also provides a measurement of the charge resolution for $N_{skip}=$\,64 through the fitted standard deviation of the pixel value distributions. A charge resolution $\sigma_{\text{e}}=$\,0.73\,\elec{} is found, notably constant up to charges of 550\,\elec{}, as shown in Figure\,\ref{fig:64_res}.

The image processing then proceeds to group individual pixels in order to reconstruct the full  energy of events within the CCD. A clustering algorithm associates contiguous pixels with charge $>$\,3.6$\sigma_{\text{e}}$ starting from a seed pixel with charge $>$\,4.6$\sigma_{\text{e}}$. These thresholds limit the number of zero-electron clusters that pass because of finite resolution to $<$\,5\%.
Defects in the CCD array result in a few pixels with defects that have charge present in a large fraction of images. A higher than average rate of clusters is then found to correspond with these regions. To avoid a bias in the energy spectrum, clusters overlapping these regions are rejected. Lastly, the cluster energy is calculated by multiplying the cluster charge in electrons by $\epsilon_{eh}=$\,3.74\,eV, the average energy required to produce an electron-hole pair in silicon at the CCD operating temperature\,\cite{2020Ramanathan}.

%% file: 4p_simulation.tex
\section{Simulations}
\label{sec:simul}
\begin{figure}[h!]
    \centering
    \includegraphics[width=\linewidth]{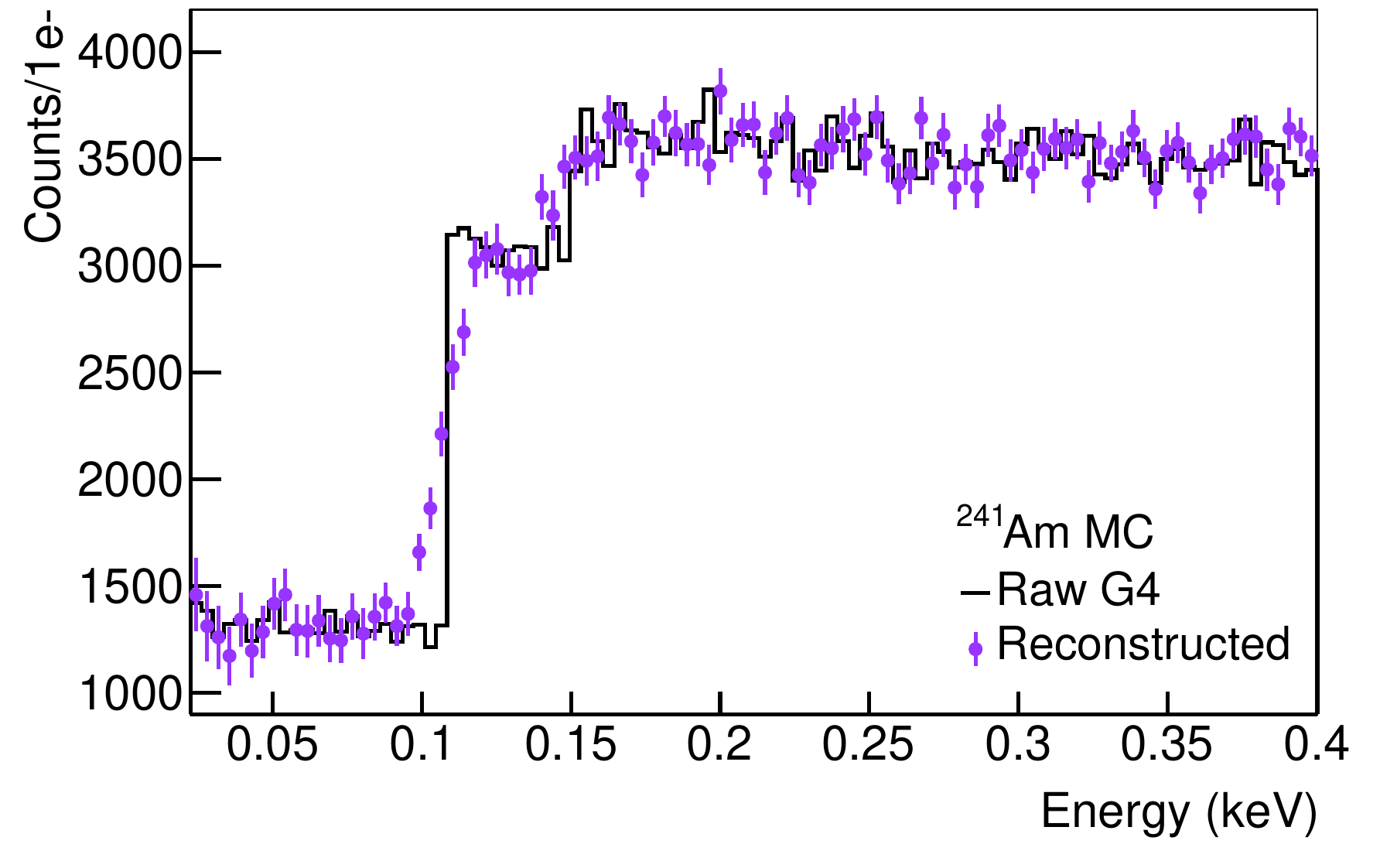}
    \caption{Reconstructed low-energy spectrum from the Monte Carlo simulation of the Compton scattering experiment (purple). The spectrum reproduces the expected features of the relativistic impulse approximation model implemented in \geant (black, generator level spectrum). The smearing in the reconstructed spectrum comes from the Fano resolution and pixel readout noise.} 
    \label{fig:MCspectrum}
\end{figure}

A full simulation of the experiment is essential to validate the data analysis methods, determine the reconstruction efficiency and interpret the results. We use the \geant simulation toolkit\,\cite{AGOSTINELLI2003250} to develop an accurate description of the geometry and materials of the experiment, including the chamber, the detector, and the \Am source. 
\geant provides the energy $E_{dep}$ deposited by particle interactions\footnote{The Livermore low-energy electromagnetic models implemented in \geant were used for the simulation. The Penelope and Monash models were also used to crosscheck the Compton spectrum.} in the silicon bulk of the CCD as well as its location (x-y-z). To convert the energy into the number of electrons, the model developed in Ref.\,\cite{2020Ramanathan} is implemented in a dedicated Monte Carlo simulation. First, for $E_{dep}>$\,50\,eV, the number of electrons is calculated by dividing $E_{dep}$ by $\epsilon_{eh}$ and smearing according to the Fano energy resolution:

\begin{equation}
\label{eq:Fano}
    \sigma_{E_{dep}}^2 = \epsilon_{eh} F E_{dep}, 
\end{equation}

\noindent where 
$F=$\,0.128 is the Fano factor measured in Ref.\,\cite{Ramanathan2017MeasurementDetector}. We note that other measurements at similar temperatures observe a value near $F=$\,0.118\,\cite{LOWE2007367,RODRIGUES2021}, however this has a negligible impact on the smeared spectrum. For $E_{dep}<$\,50\,eV, the number of electrons is obtained by sampling electron-hole pair creation probabilities\,\cite{2020Ramanathan}. Then, all resulting electrons are laterally diffused according to the parameters measured with cosmic rays tracks in the same CCD (for details of the method see\,\cite{PhysRevD.105.062003}) and distributed on the x-y pixel array. Lastly, simulated \Am clusters are pasted onto images from the background data set to properly include the pixel readout noise, the dark current, and the presence of cosmic rays and other tracks. The number of simulated clusters overlayed per image and their spatial distribution is chosen to reproduce the \Am source data resulting in a set of images which closely resembles the source data and can be processed through the same analysis chain described in Section\,\ref{sec:recon}. The reconstructed energy spectrum from these simulated images is shown in Figure\,\ref{fig:MCspectrum} between 20 and 300\,eV. It matches the expected features of the RIA model implemented in \geant, indicating that clusters are reconstructed with high efficiency and accuracy.

The reconstruction efficiency was confirmed by a dedicated Monte Carlo simulation following the steps described above. Point-like energy deposits from a uniform energy and spatial distribution were generated, diffused and pasted onto images from the background data set. The reconstructed clusters were then compared to the original energy deposits. From this study, we determine the reconstruction efficiency to be near 100\% for energy deposits as low as 15\,eV. 

%% file: 5_analysis.tex

\section{Compton spectrum measurement} 
\label{sec:analysis}
\begin{figure}[h!]
    \centering
    \includegraphics[width=\linewidth]{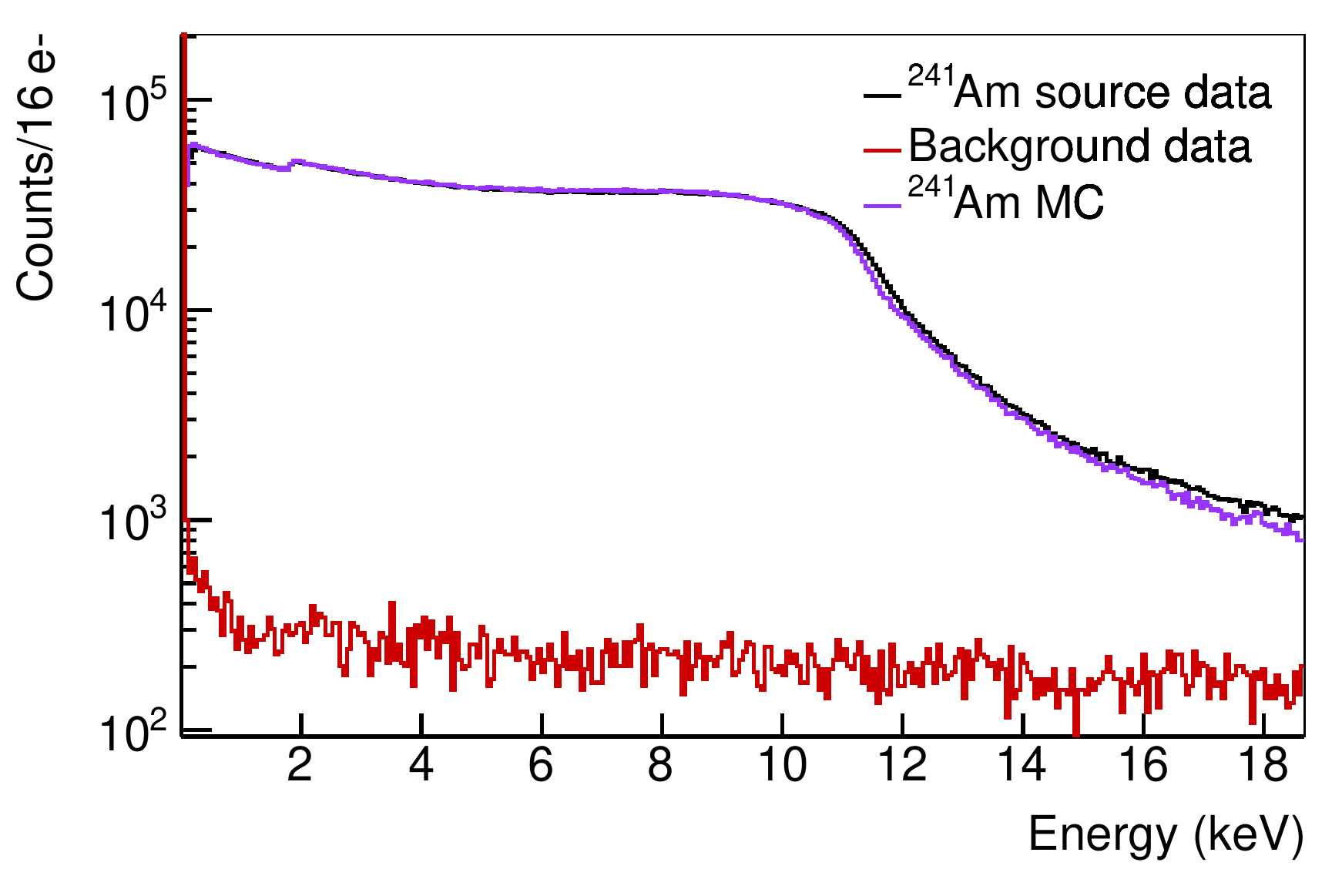}
    \caption{Measured \Am{} source spectrum (black) in $16$\,\elec{} bins. Also shown are the normalized spectra from the Monte Carlo simulation (purple) and background data (red).}  
    \label{fig:4x4AmSpectra}
\end{figure}

The energy spectrum derived by applying the analysis procedure (see Section\,\ref{sec:recon}) to the \Am{} source data is shown in Figure\,\ref{fig:4x4AmSpectra} for energies up to around 18\,keV\,\footnote{Higher energies are affected by saturation effects in the electronics, which was optimized for single-electron resolution.}. Also shown are the spectra corresponding to background data and Monte Carlo simulation normalized to exposure time of the \Am{} source data. Characteristic features of Compton scattering of the 59.54\,keV $\gamma$-ray in silicon, the Compton edge at 11.2\,keV and the K-shell step at 1.8\,keV, are evident in the measured spectrum and accurately reproduced by the Monte Carlo simulation. The slight mismatch above the edge is due to limitations in extrapolating the calibration curve, which was measured only up to 2.1\,keV, as electronics were optimized for the L-shell region (Section\,\ref{sec:recon}), to higher energies.
Also note that $\gamma$-ray lines between $10$ and $24\,\mathrm{keV}$ usually observed with \Am{} are blocked by the aluminium shield in front of the source (as described in Section\,\ref{sec:setup}). A detailed analysis of the K-shell feature is given in Section\,\ref{sec:model}. Spectra in all remaining figures are shown in 1\,electron (1\,{e}$^-$) bins to highlight the single-electron energy resolution of the measurement.

While the backgrounds are very small above 1\,keV, their contribution to the spectrum in the L-shell region cannot be neglected. The dominant source are so-called ``horizontal clusters" from the serial register. These events are common when operating on the surface, as cosmic rays and natural radiation can hit the serial register and generate charge. As the rows are read out, the clusters are reconstructed as horizontal tracks in the CCD's active area. Since the full energy of the particle is not deposited, the horizontal clusters contain only a few pixels and cannot be distinguished from real low energy events. Thus, a background subtraction procedure is required to reach the lowest threshold. Since the rate of horizontal clusters is proportional to the flux of radiation incident on the CCD, their number should increase significantly in presence of the \Am source. This was verified with a dedicated data set. With the source in place, images were taken by moving charge towards the CCD active area, in the opposite direction of the readout serial register. This operating mode results in images containing only clusters originating in the serial register. We confirmed that the rate of horizontal clusters with the \Am source is a factor of ten higher than the rate of clusters measured in the standard background runs. Therefore, both the serial register and standard background data must be considered. Given the relevance of this effect below 400\,eV, two independent methods were developed for an accurate measurement of the \Am Compton spectrum.

\begin{figure}[h!]
    \centering
    \includegraphics[width=\linewidth]{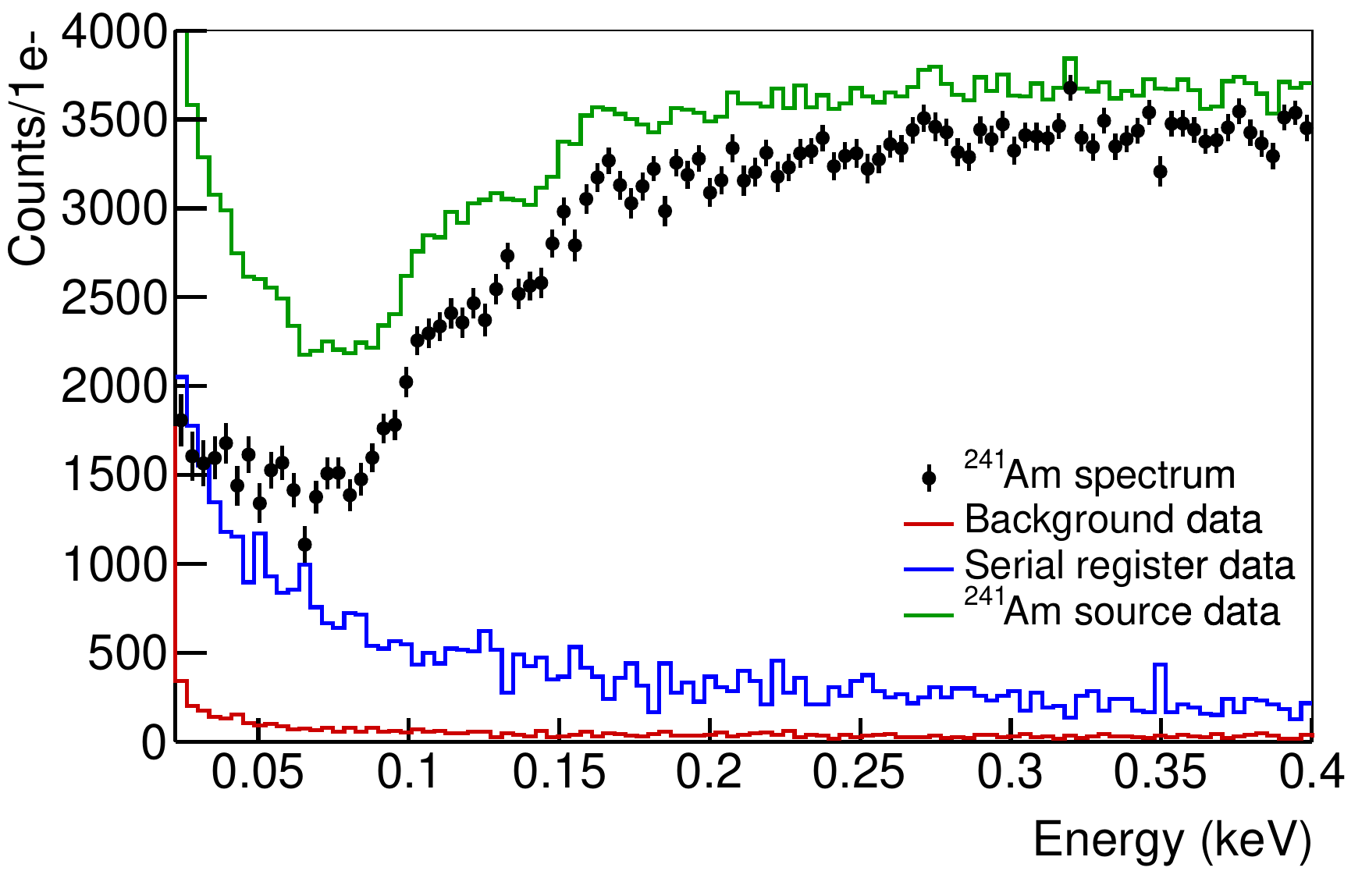}
    \caption{The subtracted \Am Compton spectrum (black) in the L-shell region. To illustrate the measurement components, the normalized spectra used in the subtraction are shown: \Am source (green), \Am serial register (blue), and background (red).}  
    \label{fig:Lstep_sub}
\end{figure}

In the first method, we perform a bin-by-bin subtraction of the serial register background and standard background spectra from the \Am source spectrum. The subtracted spectra were normalized to the exposure time of the source data. This approach accounts for both the increased rate of horizontal clusters due to the source (serial register background) and background clusters in the CCD active area from cosmic rays and radiogenic sources in the apparatus (standard background). The spectral subtraction method is illustrated in Figure\,\ref{fig:Lstep_sub}, where all components and the derived Compton spectrum are shown. 

\begin{figure}[h!]
    \centering
    \includegraphics[width=\linewidth]{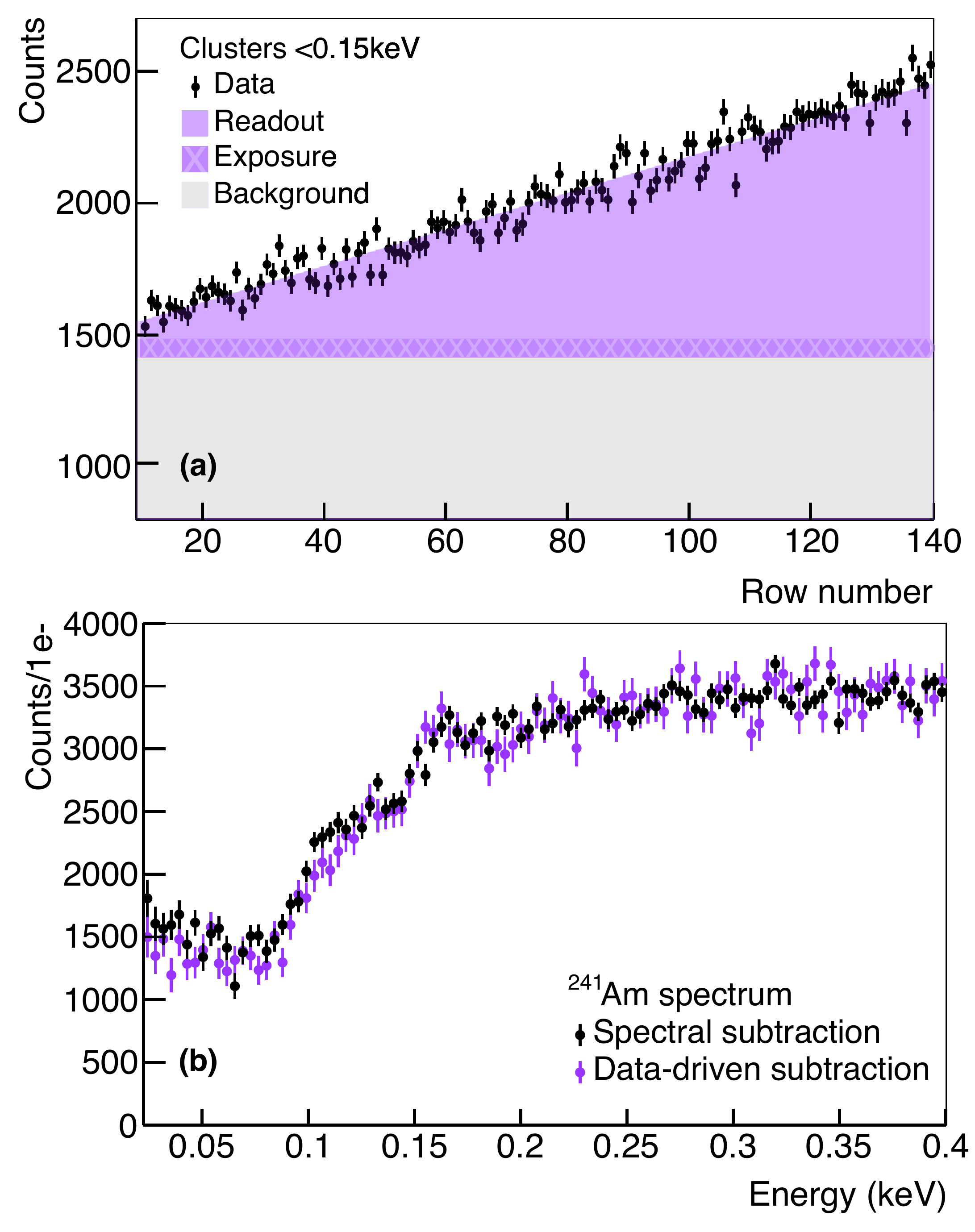}
    \caption{(a) Number of reconstructed clusters below 0.15\,keV as a function of row in the CCD. The area of the  triangle (purple) is taken as an estimate of the signal, together with the small rectangle right below (purple crosses) which accounts for 3\,s exposure before readout. The larger rectangle (gray) represents the subtracted background.
    (b) Comparison of the spectral subtraction (black) and the data-driven (purple) spectrum in the L-shell energy region.}
    \label{fig:area_method}
\end{figure}

In the second method, a data-driven approach uses only the \Am source data to estimate the background, as illustrated in Figure\,\ref{fig:area_method}. 
Due to the sequential readout of the CCD, the effective exposure time of a row increases linearly with its readout order. The number of clusters produced by Compton scattering in the active area increases as a function of row position in the image. However, the number of horizontal clusters should remain constant since the exposure time in the serial register is the same for all rows. We thus perform a linear fit of the number of clusters as a function of row to estimate the signal, which is represented in the area of the triangle illustrated in Figure\,\ref{fig:area_method}{\color{red}(a)}). A small correction ($\approx$10\%) is applied to take into account that the CCD is exposed for 3\,s (during which clusters accumulate uniformly over the rows) and then read out for 37\,s. The procedure was performed on each energy bin, with size 3.74\,eV\,(1\,\elec{}), to produce the final low energy spectrum shown in Figure\,\ref{fig:area_method}{\color{red}(b)}). The derived spectrum is compared to the one obtained with the spectral subtraction method. The two spectra are in excellent agreement showing similar features. It should be noted that the spectra are not normalized to each other; their agreement in the absolute rate indicates that the same amount of signal is recovered by both methods.

Several checks were performed to validate the analysis procedure. The data-driven method was applied to the serial register background data set, verifying that the rate of horizontal clusters indeed does not depend on the row number. The serial register background spectrum was found to closely match that obtained from the overscan portion of the \Am source images, which also contains only clusters originating in the serial register. Monte Carlo simulated clusters were pasted onto images of the serial register background and standard background data sets, as explained in Section\,\ref{sec:simul}. Furthermore, \Am source data were divided into two chronological, independent sets, and the corresponding spectra were found to be compatible within statistical uncertainty. The analysis was also repeated with lower thresholds in the clustering algorithm (seed pixel threshold $>$\,4\,$\sigma_{\text{e}}$ and contiguous pixel threshold $>$3\,$\sigma_{\text{e}}$), resulting in a very consistent spectrum. All these checks give confidence that the measured spectrum is accurate down to threshold with uncertainties dominated by statistics.

%% file: 6_model.tex

\section{Comparison with theoretical models} 
\label{sec:model}

An accurate estimate of low energy backgrounds from Compton scattering of radiogenic $\gamma$-rays is important for DAMIC-M and other future direct detection experiments with energy thresholds of $\sim$10\,eV\,\cite{SENSEI:2020dpa,SuperCDMS:2020ymb,Aguilar-Arevalo:2022kqd,CRESST:2020wtj}. The Monte Carlo packages used by these experiments, e.g. \geant\,\cite{AGOSTINELLI2003250} and MCNP\,\cite{mcnp}, employ the RIA model to simulate Compton scattering (see Section\,\ref{sec:compton}). We use the measured spectrum obtained from the spectral subtraction method (first method) to test the validity of the RIA model in an unexplored energy range, and thus the appropriateness of these Monte Carlo simulations, down to 23\,eV. Note that the assumptions of the RIA model are not valid for energy transfers close to the electron binding energy and we may expect its predictions to be inaccurate near the silicon steps. Having already shown the agreement with the \geant simulation at the Compton edge in Figure\,\ref{fig:4x4AmSpectra}, here we compare the model to data near the atomic shells.

\begin{figure*}[]
    \centering
    \includegraphics[width=1.\textwidth]{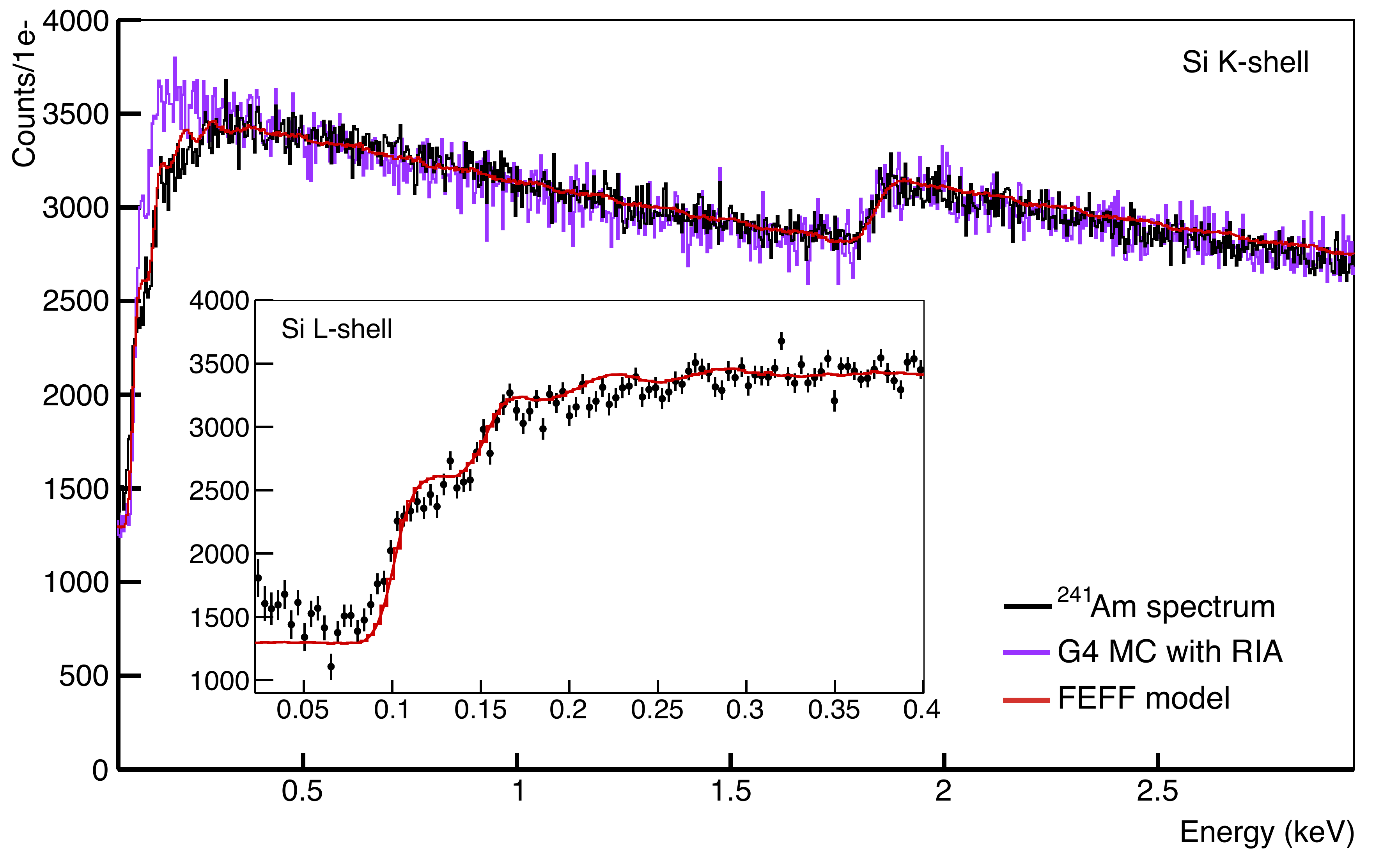}
    \caption{The measured \Am Compton spectrum (black) from the 23\,eV detection threshold to $2.1\,\mathrm{keV}$. The K-step is observed at 1.8\,keV. The \geant simulated spectrum (purple) that is based on the relativistic impulse approximation is also shown. In red is the {\it ab initio} calculation from the FEFF code, with detector response taken into account. The inset shows the data comparison to the FEFF prediction in the L-shell energy range.} 
    \label{fig:kshell}
\end{figure*}

Previous measurements of the K-shell region have been in good agreement with RIA model\,\cite{Ramanathan2017MeasurementDetector}. Our skipper CCD measurement is in alignment with these results, as shown in Figure\,\ref{fig:kshell}. The simulated \geant Compton spectrum, which incorporates both the RIA model and silicon detector response, provides a close match to the data down to about 0.5\,keV. In particular, the K-shell transition step and slopes of the spectrum before and after the K-shell energy are reproduced. 

However, there are notable differences at lower energies, as shown in  Figure\,\ref{fig:kshell} and its inset. A softening of the spectrum below 0.5\,keV is observed in the data, confirming the previous measurement\,\cite{Ramanathan2017MeasurementDetector}. Data between the $\rm{L_{2,3}}$ and $\rm{L_{1}}$ energies are compatible with a step, a feature predicted by the RIA model, but has a softer shape.
We detect for the first time a plateau below the $\rm{L_{2,3}}$ energy ($99.2\,\mathrm{eV}$) corresponding to Compton scattering on valence electrons. Its measured amplitude is consistent with the expectation of scaling by the number of electrons available in the shell\footnote{In a recently reported measurement also with a skipper CCD\,\cite{Botti:2022lkm} this expectation was not verified, due to the presence of unsubtracted backgrounds as explicitly stated in the reference.}. 
Overall the \geant Monte Carlo overestimates the measured spectrum by up to 20\% below 0.5\,keV and in the L-shell region.

We then compared the data with {\it ab initio} calculations from the FEFF code (see Section\,\ref{sec:compton}).  
The FEFF predictions were obtained by computing the corresponding $S_{nl}(q,E)$ (see Equation\,\ref{eq:genxsection}) in discrete $q(\cos \theta)$ and $E$ steps and summing over all scattering angles. The code used was FEFF10 with configuration and silicon crystal structure from the Materials Project\,\cite{materialsproject} (Materials Id mp-149).
The computation was performed in real-space (vs. momentum-space) on a cluster of 35 atoms centered on the target atom. For the L- and K-steps, for each $q(\cos \theta)$ and $E$ step, the code was executed twice, once with the {\tt XANES} card to obtain the fine structure near the atomic edge, and once with the {\tt EXAFS} card to obtain the extended energy region. To compute the underlying spectrum from valence electrons, the code was executed with the {\tt COMPTON} card, whose accuracy in silicon has been previously validated with data\,\cite{FEFFCompton}. Note that the plasmon and particle-hole pair excitations, which contribute to $S_{nl}(q,E)$ at very low momentum transfers\,\cite{SternemannPlasmon}, are not included.
We checked for convergence and numerical consistency of the FEFF output.
The original configuration option for the interaction with the vacated core hole was {\tt Corehole Rpa}, an  approximation similar to that in the Bethe-Salpeter equation (BSE)\,\cite{BSE}, and resulted in an L-step that was too sharp. Thus, it was consequently modified to {\tt Corehole none} for the L-shell, which provides a much better match to the data.This is consistent with previous  NRIXS measurements in silicon, where the omission in the calculation of the core-hole interaction with the photoelectron was observed to better match the spectrum in the extended energy region\,\cite{Sternemann1,Sternemann2}.
For a realistic comparison of the FEFF model, the spectrum was convoluted with the detector response of the silicon CCD, including the experimental charge resolution and Fano resolution. The resulting FEFF prediction for the Compton spectrum is shown in Figure\,\ref{fig:kshell} and its inset. There is an excellent agreement with the data over the entire energy range. In particular, FEFF reproduces L-shell features to better than 10\% , where a softening of the spectrum and step is observed. The RIA model fails in this region.

Using the FEFF model as a reference, we can gain useful insight on the detector response to ionization in silicon. The FEFF model reproduces the location and relative height of the K-shell step as well as the slope of the spectrum before and after the step. Note that the energy scale of the measured spectrum is determined by the value of $\epsilon_{eh}$ used to convert the charge into energy (3.74\,eV per electron, see Section\,\ref{sec:recon}). We performed a fit of the data with the FEFF model keeping the K-step location as a free parameter, deriving

\begin{equation}
    \epsilon_{eh}(1839\,\mathrm{eV})=3.755\pm0.008_{stat}\pm0.010_{syst}\,\mathrm{eV}, 
\end{equation}

\noindent where the systematic uncertainty takes into account the fitting range, the energy scale calibration, the CCD temperature, and the choice of the model. The result is in excellent agreement with the nominal value of $\epsilon_{eh}$ and confirms the accuracy of the calibration detailed in Section\,\ref{sec:recon}.
A similar procedure could be applied at the L-shell energies but there are theoretical uncertainties in the {\it ab initio} calculations \,\cite{privatec} and more statistics are required to validate the predicted structures. This makes it difficult to derive a meaningful value of the Fano factor since it strongly depends on the detailed shape of the predicted spectrum. Nevertheless, the excellent match of the model to data is consistent within few percent with a constant value of $\epsilon_{eh}$ down to 100\,eV, in agreement with previous measurements\,\cite{Scholze98}. Moreover, it has been proposed that energy loss mechanisms may vary in this energy range, resulting in an energy-dependent Fano factor and $\epsilon_{eh}$\,\cite{FRASER1994368,osti_913319}. We tested this hypothesis by convolving the FEFF prediction with the energy-dependent silicon detector response model parameters from Ref.\cite{osti_913319} and the resulting spectrum provided a significantly worse fit of the data, confirming that an energy-independent detector response model is adequate down to low energy.

%% file: 7_conclusion.tex

\section{Discussion}
\label{sec:conclusions}

Future dark matter experiments require a robust knowledge of backgrounds from Compton-scattered environmental $\gamma$-rays down to the eV-scale. To estimate these backgrounds, the community relies heavily on Monte Carlo packages such as \geant, which incorporates the relativistic impulse approximation model to describe the scattering physics. However, it is known that this model does not to apply to small scattering angles, and previous CCD measurements in silicon have observed a softer L-shell step than predicted. Thus, precision measurements of the Compton spectrum at low energies are needed to determine if the discrepancies are due to unknown detector effects or our knowledge of the cross sections themselves.

Our measurement explores the Compton spectrum in silicon down to a threshold of $23\,\mathrm{eV}$, made possible with sub-electron resolution of a DAMIC-M prototype CCD. We detect for the first time Compton scattering on valence electrons below 100\,eV, and clearly identify features associated with the silicon K, L$_{1}$, and L$_{2,3}$-shell. The RIA model expectations are in very good agreement with data above 0.5\,keV, but fail to reproduce the spectrum in the L-shell region and overestimate rates by up to 20\%. Since RIA-based simulations are used to build background models for direct detection experiments, care should be taken to evaluate its impact on the sensitivity to a potential dark matter signal. It may also be necessary to improve the Compton scattering model at low energy in Monte Carlo codes, such as \geant. In this respect, we have found our measured spectrum to be in much better agreement with predictions from {\it ab initio} calculations of Compton scattering with bound electrons using the FEFF code. To our knowledge, this is the first time such calculations, usually employed to evaluate X-ray scattering data at fixed momentum transfers, are compared to a Compton spectrum. The FEFF model predicts features in the spectrum that will require additional statistics to be confirmed. Data reported here are consistent with an energy-independent value of $\epsilon_{eh}$, the average energy to produce an electron-hole pair in silicon. For a meaningful test of the energy dependence of the Fano factor, which describes the fluctuations in the energy deposited through scattering, a more precise measurement is required as well as a solid theoretical prediction of the spectrum shape. For this purpose we plan to take additional measurements to validate the model at different $\gamma$-ray energies using a $^{57}$Co source ($E_{\gamma}=$\,122 and 136\,keV). We will also perform estimation studies to quantify the impact the FEFF model has on the DAMIC-M low-mass sensitivity.

Lastly, we note that for this measurement a skipper CCD was operated continuously for several months with excellent stability. Calibration and event reconstruction procedures optimized for sub-electron resolution were developed. Our precise determination of the Compton spectrum demonstrates that skipper CCDs developed for DAMIC-M detect with high efficiency and accuracy energy deposits of just a few ionization charges in the silicon bulk. These results give us confidence in the forthcoming search for dark matter particles with the DAMIC-M experiment.

%% file: 8_acknowledgements.tex

\section{Acknowledgements}
The DAMIC-M project has received funding from the European Research Council (ERC) under the European Union's Horizon 2020 research and innovation programme Grant Agreement No. 788137, and from NSF through Grant No. NSF PHY-1812654. 
The work at University of Chicago and University of Washington was supported through Grant No. NSF PHY-2110585. This work was supported by the Kavli Institute for Cosmological Physics at the University of Chicago through an endowment from the Kavli Foundation. 
We also thank the College of Arts and Sciences at UW for contributing the first CCDs to the DAMIC-M project.
IFCA was supported by project PID2019-109829GB-I00 funded by MCIN/ AEI /10.13039/501100011033.
The Centro At\'{o}mico Bariloche group is supported by ANPCyT grant PICT-2018-03069.
The University of Z\"{u}rich was supported by the Swiss National Science Foundation.
The CCD development work at Lawrence Berkeley National Laboratory MicroSystems Lab was supported in part by the Director, Office of Science, of the U.S. Department of Energy under Contract No. DE-AC02-05CH11231.


We thank Gerald T.\,Seidler for introducing us to the FEFF code, and to Joshua\,J. Kas, Micah\,P. Prange, and John\,J. Rehr for their support with FEFF. We also thank Christian Sternemann for sharing his NRIXS silicon spectra.